\documentclass[twocolumn,showpacs,amsmath,amssymb,pre]{revtex4}

\usepackage{graphicx}
\usepackage{color}
\usepackage{dcolumn}
\usepackage{amsmath}
\usepackage{CJK}
\usepackage{subfigure}
\usepackage{sidecap}

\usepackage{mathrsfs}
\usepackage{amsfonts}
\usepackage{indentfirst}

\newcommand{\be}{\begin{equation}}
\newcommand{\ee}{\end{equation}}
\newcommand{\bey}{\begin{eqnarray}}
\newcommand{\eey}{\end{eqnarray}}
\newcommand{\bw}{\begin{widetext}}
\newcommand{\ew}{\end{widetext}}

\newcommand{\ra}{\rangle}
\newcommand{\la}{\langle}

\newcommand{\ba}{\begin{array}}
\newcommand{\ea}{\end{array}}
\newcommand{\bi}{\begin{itemize}}
\newcommand{\ei}{\end{itemize}}
\newcommand{\bem}{\begin{enumerate}}
\newcommand{\eem}{\end{enumerate}}

\begin{document}

\title{Probing the excited-state quantum phase transition through statistics of Loschmidt echo and quantum work}

\author{Qian Wang$^{1,2}$ and H.~T.~Quan$^{1,3}$\footnote{ Electronic address: htquan@pku.edu.cn}}

\affiliation{$^{1} $School of Physics, Peking University, Beijing 100871, China\\
 $^{2}$Department of Physics, Shanghai Normal University, Shanghai 200234, China  \\
 $^2$Collaborative Innovation Center of Quantum Matter, Beijing 100871, China}

\date{\today}

\begin{abstract}
  By analyzing the probability distributions of the Loschmidt echo (LE) and quantum work,
  we examine the nonequilibrium effects of a quantum many-body system, which exhibits an excited-state quantum
  phase transition (ESQPT).
  We find that depending on the value of the controlling parameter the distribution of the LE displays different patterns.
  At the critical point of the ESQPT, both the averaged LE and the averaged work show a cusplike shape.
  Furthermore, by employing the finite-size scaling analysis of the averaged work, we obtain the critical exponent of the ESQPT.
  Finally, we show that at the critical point of ESQPT the eigenstate
  is a highly localized state,
  further highlighting the influence of the ESQPT on the properties of the many-body system.
\end{abstract}


\maketitle

\section{Introduction}

 Quantum phase transition (QPT) \cite{sachdev,vojta,dutta} is characterized by a dramatic change of the ground-state properties
 of a quantum system when the controlling parameter, such as an external magnetic field or an internal coupling
 strength \cite{suzuki}, passes through the critical point.
 It occurs at absolute zero temperature and the change of phases is solely driven by quantum fluctuations.
 In the past few decades, QPT has become a vast interesting topic and has attracted lots of attention.
 It is well known that QPT has strong influence on the property of the system, such as the divergence of the derivative of the
 ground-state entanglement (concurrence) \cite{amico}, fidelity \cite{zanardi}, and the emergence of chaos \cite{emary}.
 The effects of the QPT on the nonequilibrium properties of the system also attracted lots of
 attention \cite{dutta2,haitao,wyr,rossini,paz,zurek,polk,polkovnikov,agsilva,dgcm}.
 In particular, QPT has been observed in many experiments \cite{suter,ditty,greiner,bloch,baumann}.

 Recently, the quantum critical phenomenon has been extended to the excited states of the system \cite{macek,caprio,psmpc,ribeiro2}.
 An excited-state quantum phase transition (ESQPT) refers either a nonanalytic variation of eigenenergies of the individual
 excited states with respect to the controlling parameter
 or a singular behavior of the density of states \cite{caprio,psmpc}.
 They appear in various models, such as Lipkin-Meshkov-Glick (LMG) model \cite{ribeiro1,ribeiro2,relano,perez,zigang,cejnar,santos},
 molecular vibron model and interacting boson model (IBM) \cite{pcrf,caprio},
 Dicke and Jaynes-Cumming models \cite{brandes,mabm},
 Rabi model \cite{puebla2}, kicked top \cite{bastidas}, and microwave Dirac billiards \cite{dietz}.
 Moreover, the signatures of ESQPTs have been observed in many experiments \cite{dietz,bpw,zobov,larese,larese2,larese3,dietz2}.
 Here, it is worth pointing out that even though there are connections between excited energy and the temperature in an isolated system,
 ESQPTs and thermal phase transitions are qualitatively different \cite{ardp}.
 The static definition of the order parameter is invalid for ESQPTs.
 To define the order parameter of an ESQPT, one must take into account
 some dynamic properties of the system \cite{caprio,puebla}.

 The influence of ESQPTs on the dynamics of the system has been studied in literature.
 Various dynamic consequences have been predicted, such as the enhancement of the decoherence in open systems \cite{relano,perez},
 the peculiar behavior of the survival probability in isolated systems \cite{santos},
 the emergence of symmetry-breaking steady states \cite{puebla},
 the singularities in the evolution of observables \cite{gevm,wktb},
 and the abrupt increase of entropy \cite{lobez}.
 To get a better understanding of the ESQPTs,
 however, more works are required to study the nonequilibrium dynamic effects of the system, which exhibits an ESQPT.


 In the current study, we analyze the nonequilibrium dynamics of an isolated quantum many-body system that undergoes an ESQPT.
 Specifically, we study the dynamics following a sudden quench of the controlling parameter, i.e., an external magnetic field, in the LMG model.
 The aim of this work is to investigate how the signatures of an ESQPT manifest themselves in the nonequilibrium dynamics when the controlling
 parameter is quenched across the critical point.
 By studying the probability distribution of the Loschmidt echo (LE), we show the effect of ESQPT on the dynamics in the LMG model.
 From the quantum thermodynamics perspective, we study the statistics of work done on the LMG model in the sudden quench process.
 We find that the signature of the ESQPT can be observed in both the average value of work and the standard deviation of work distribution.
 By applying the finite-size scaling analysis, we obtain the critical exponent of the ESQPT.
 In order to better understand the influence of ESQPT on the dynamics of the system, we finally investigate the spectral function.
 Our results show that the eigenstate of the system is a highly localized state at the critical point of the ESQPT.

 Here we should point out that the effects of the ESQPT on the dynamics in the LMG model have been studied previously.
 In Refs.~\cite{relano,perez}, the decoherence of the central spin system, which is affected by the ESQPT, was explored.
 The time evolution of the LE with the
 initial state given by the eigenstates of $\mathrm{U}(1)$ and $\mathrm{SO}(2)$ part of a
 Hamiltonian was examined in Refs.~\cite{santos}.
 However, our study is different from those works in that we seek the connection between the
 statistics of the LE and the ESQPT.
 In addition, we study the effects of the ESQPT on the thermodynamic properties of the system.
 To achieve these goals, we set the initial state, denoted as $|\mathrm{ini}=n\ra$, to
 be the $n$th eigenstate of the initial Hamiltonian.
 Also, we fix the amplitude of quench in our study.

 The remainder of this article is organized as follows.
 In Sec.~\ref{model}, we briefly review the properties of the LMG model and show that the ESQPT occurs
 either when varying the controlling parameter or when varying the energy (quantum number).
 In Sec.~\ref{results}, we study in detail the statistical properties of LE and quantum work.
 We show that both two quantities can be used to characterize the ESQPT.
 Section~\ref{sfl} provides the analysis of the spectral function.
 We find the onset of localization in the eigenstate.
 Finally, we give our summary and discuss our results in Sec.~\ref{sum}.

 \begin{figure}
  \includegraphics[width=\columnwidth]{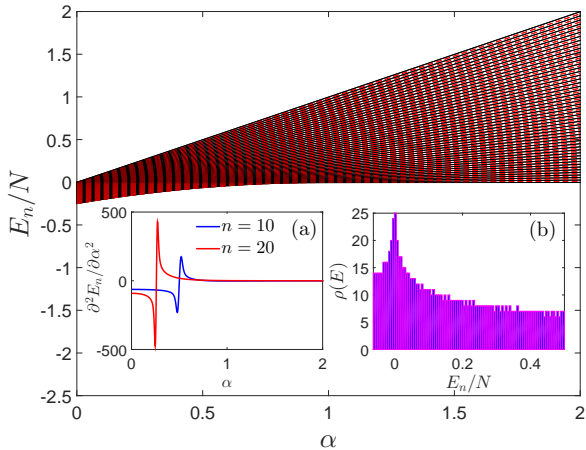}  
  \caption{(Color online)
   The energy levels of $\mathcal{H}$ as a function of $\alpha$ for $N=100$.
   The black solid curves denote the even-parity levels, while the odd-parity levels are denoted by the red dashed curves.
   Inset: (a) the second derivative of eigenenergy with respect to $\alpha$ for different eigenstates $|n\ra$,
   with $n=10$ (right peak) and $20$ (left peak);
   (b) the density of states for
    $\mathcal{H}$ with $\alpha=0.5$ and $N=1000$.}
  \label{energyC}
 \end{figure}

\section{The LMG Moedl} \label{model}

 We study the so-called Lipkin-Meshkov-Glick (LMG) model, which describes a set of $N$ spin-$1/2$
 mutually interacting with each other in a transverse field. The Hamiltonian of the LMG model reads
 \cite{Lipkin,dusuel2,botet1}
 \be
    \mathcal{H}=-\frac{1}{N}\sum_{i<j}\left(\sigma_x^i\sigma_x^j+\gamma\sigma_y^i\sigma_y^j\right)+\alpha\sum_i\sigma_z^i,\label{full}
 \ee
 where $\sigma_{x,y,z}^i$ are the Pauli matrices on the $i$th site in the spin chain,
 $\alpha$ denotes the strength of the magnetic field, and $\gamma$ is the anisotropy parameter.
 For the sake of simplicity, we set $\gamma=0$ in our study.  
 Here, we should point out that for $\gamma\neq0$ case, the full LMG model (\ref{full}) is expected to exhibit more complex
 phase diagram \cite{ribeiro1,ribeiro2} than the case we studying in this paper. However, the results and conclusions in our paper
 do not change for the more general case.
 Using the total spin operator $S_\beta=\sum_{i=1}^N\sigma_\beta^i$ with $\beta=\{x,y,z\}$, the Hamiltonian
 can be rewritten as \cite{relano,perez,zigang,santos}
 \be \label{LPH}
   \mathcal{H}=-\frac{1}{N}(S_x)^2+\alpha\left(S_z+\frac{N}{2}\right).
 \ee

 One can rewrite the Hamiltonian (\ref{LPH}) into a two-level bosonic Hamiltonian via the Schwinger
 transformation \cite{perez,zigang},
 \be
    S_+=t^\dag s=(S_-)^\dag,\
    S_z=\frac{1}{2}(t^\dag t-s^\dag s),
 \ee
 where $S_{\pm}=S_x\pm iS_y$ and $s^\dag, t^\dag$ are the creation operators of two species scalar bosons $s$ and $t$, respectively.
 Finally, the Hamiltonian (\ref{LPH}) becomes
 \be \label{BH}
    \mathcal{H}=\alpha \hat{n}_t-\frac{1}{4N}\hat{Q}^2,
 \ee
 where $\hat{n}_t=t^\dag t$, $\hat{Q}=t^\dag s+s^\dag t$.
 Obviously, the total number of bosons is a conserved quantity, i.e.,
 $[\hat{N},\mathcal{H}]=0$ with $\hat{N}=t^\dag t+s^\dag s$.
 We can write down the Hamiltonian (\ref{BH}) in the following bases
 \be
    |N,n_t\rangle=\frac{(\hat{t}^\dag)^{n_t}(\hat{s}^\dag)^{N-n_t}}{\sqrt{(n_t)!(N-n_t)!}}|0\rangle,
 \ee
 where $N\geq n_t\geq0$ and $|0\rangle$ is the vacuum state \cite{perez,zigang,santos}.
 Then the dimension of the Hilbert space is $\mathrm{Dim}[\mathcal{H}]=N+1$.
 The nonzero elements of the Hamiltonian matrix read as
 \bey
   \langle N,n_t|\mathcal{H}|N,n_t\rangle&=&\alpha n_t+f, \\
   \langle N,n_t|\mathcal{H}|N,n_t+2\rangle&=&-\frac{1}{4N}\sqrt{(n_t+1)(N-n_t)} \nonumber \\
                                            &&\times\sqrt{(n_t+2)(N-n_t-1)},
 \eey
 where $f=-1/(4N)[(n_t+1)(N-n_t)+n_t(N-n_t+1)]$.
 One can obtain the eigenstates and eigenenergies through numerically diagonalizing the Hamiltonian.
 Hamiltonian (\ref{LPH}) conserves parity $(-1)^{n_t}$; therefore, the eigenstates can be labeled
 as even- or odd-parity eigenstates \cite{zigang,cejnar,santos}.
 The Hamiltonian matrix is split in two blocks, one of dimension $\mathrm{Dim}[\mathcal{H}]_{\mathrm{even}}=N/2+1$,
 and the other of dimension $\mathrm{Dim}[\mathcal{H}]_{\mathrm{odd}}=N/2$ \cite{santos}.

 It is well known that, in the thermodynamic limit, i.e., $N\to\infty$, the LMG model undergoes a second-order quantum phase transition (QPT)
 when the controlling parameter is varied across $\alpha_c=1$ \cite{Lipkin,dusuel2,botet1,htzc,qianwang}.
 At the critical point, the properties of the ground-state changes dramatically, and the second derivative of the ground-state energy
 with respect to $\alpha$ shows a discontinuity or diverges.
 The energy gap between the ground state and the first excited
 state vanishes as $\Delta_0\propto|\alpha-\alpha_c|^\nu$, with the critical exponent $\nu=1/2$ \cite{dusuel2,botet1}.
 Interestingly, these properties of the ground-state QPT also manifest themselves in the excited states.

 \begin{figure}
    \centering
    \includegraphics[width=\columnwidth]{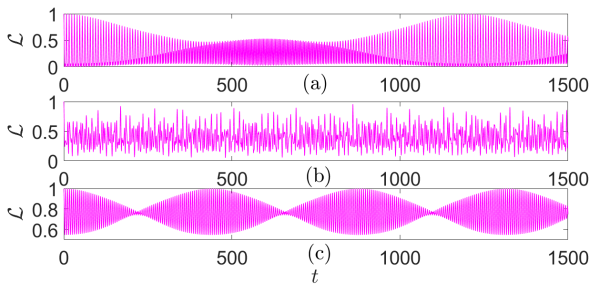}
    \caption{(Color online) Time evolution of the LE for different controlling parameters:
   (a) $\alpha=0.1$, (b) $\alpha=0.48$, and (c) $\alpha=0.8$. Here, $N=400$ and $\delta\alpha=0.01$.
   The initial state is the $77$th eigenstate of $\mathcal{H}_0$.}
   \label{TLE1}
 \end{figure}

 In Fig.~\ref{energyC}, we plot the energy levels as a function of $\alpha$ of the LMG model.
 Two remarkable features can be found from this figure.
 First, for $E<0$, the eigenstates are doubly degenerate, and nondegenerate when $E>0$.
 Moreover, each eigenvalue $E(\alpha)$ undergoes an inflection around $E\approx0$ \cite{caprio}.
 Second, when $\alpha<1$, the energy levels concentrate around $E\approx0$. The energy gap between adjacent
 energy levels is close to zero around $E\approx0$.

 According to these features, the ESQPT can be characterized by two different ways.
 On the one hand, the ESQPT can be defined by varying controlling parameter $\alpha$.
 Namely, for a specific excite state, ESQPT occurs when $\partial^2 E_n(\alpha)/\partial\alpha^2$ shows a divergence [see Fig.~\ref{energyC}(a)].
 This definition implies that different eigenstates have different critical value of $\alpha$.
 On the other hand, one can fixed the controlling parameter $\alpha$ and define the ESQPT as a singular behavior of the density of states.
 As shown in Fig.~\ref{energyC}(b), for finite $N$, the density of states displays a peak around $E=0$.
 While for $N\to\infty$, this will lead to a logarithmic divergence \cite{ribeiro2}.
 The critical energy of the LMG model (\ref{LPH}) is, therefore, $E_c=0$.
 Here, we should point out that the critical energy of ESQPT usually varies with the controlling parameter \cite{santos},
 and different eigenstates will have different critical energy.
 However, the critical energy of the LMG model (\ref{LPH}) is independent of $\alpha$ and eigenstate number.
 Although these two different definitions seem unrelated, the relation between them has been studied in Refs.~\cite{caprio} and \cite{cejnar}. 
 In the following, for both of these definitions, the influence of the ESQPT on the
 nonequilibrium dynamics will be studied.

 \begin{figure}
  \includegraphics[width=\columnwidth]{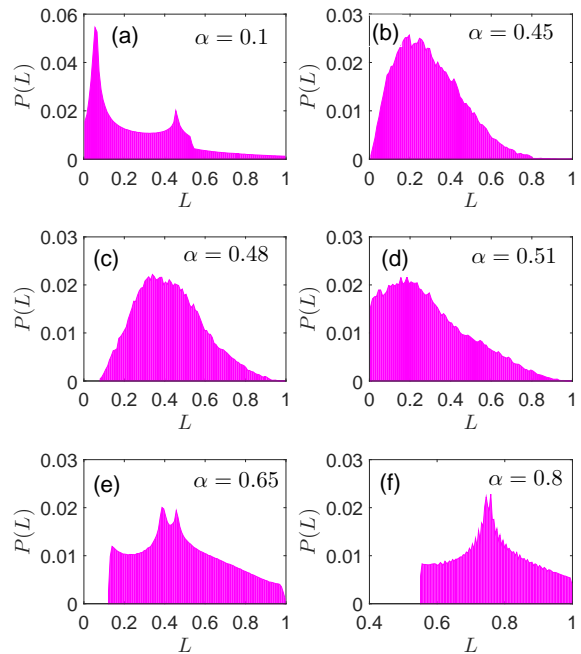}
  \caption{(Color online)
   The probability distribution of the LE for different controlling parameters with $N=400$ and $\delta\alpha=0.01$.
   The initial state is the $77$th eigenstate of $\mathcal{H}_0$.}
  \label{LED}
 \end{figure}

 \section{Nonequilibrium dynamics} \label{results}

 In this work, we consider a sudden quench process.
 That is, the magnetic field strength is changed suddenly at time $t=0$.
 The initial state is chosen to be the $n$th eigenstate of the system, i.e., $|\mathrm{ini}=n\ra$, with $\alpha_0=\alpha$.
 The initial Hamiltonian is $\mathcal{H}_0=\mathcal{H}(\alpha)$.
 Then at $t=0$ we make a sudden quench $\alpha_0\to\alpha_f=\alpha+\delta\alpha$, where $\delta\alpha$ denotes
 the amplitude of the quench.
 We study the time evolution of the system under the final Hamiltonian $\mathcal{H}_f=\mathcal{H}(\alpha_f)$,
 $|\psi(t)\rangle=\exp(-i\mathcal{H}_f t)|n\rangle$.
 The central quantity that we study is the time-dependent overlap
 \bey
    \mathcal{O}(t)=\langle n|\exp(i\mathcal{H}_0 t)\exp(-i\mathcal{H}_f t)|n\rangle
            =e^{iE^0_nt}\langle n|\psi(t)\rangle,  \label{TOL}
 \eey
 where $E_n^0$ is the eigenenergy of $|n\ra$.
 From this quantity, we can get several important signatures that can be used to probe the ESQPT.

 In this section, we mainly study two important quantities, namely, the Loschmidt echo (LE) and the
 quantum work distribution. Both of them can be derived from the time evolution of the overlap (\ref{TOL}) and have been
 widely used in many fields (e.g., see Refs.~\cite{htzc,qianwang,jalabert,gorin,polkovnikov,fusco,dasone,dastwo,halimeh,jafari} and references therein).
 In particular, it has already been demonstrated that both of these quantities can be used as
 the detector of the ground-state QPTs \cite{haitao,silva}.

 \begin{figure}
    \centering
    \includegraphics[width=\columnwidth]{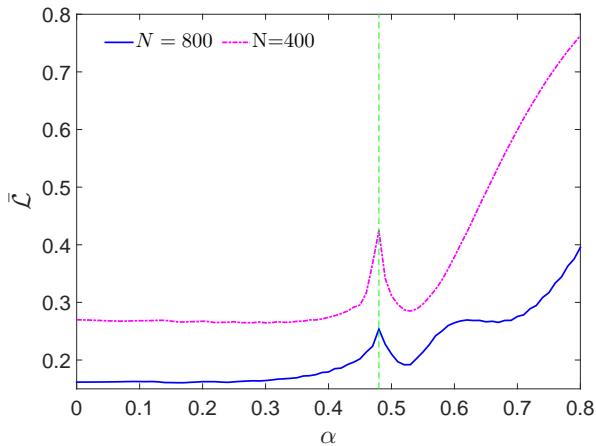}
  \caption{(Color online)
   $\bar{\mathcal{L}}$ as a function of $\alpha$ for different $N$ with $\delta\alpha=0.01$.
   The initial states are: $|\mathrm{ini}=77\ra$ for $N=400$, while $|\mathrm{ini}=157\rangle$ for $N=800$.
   The vertical green dashed lines indicate the position of $\alpha=0.48$.}
  \label{average}
 \end{figure}

 \subsection{Statistics of the Loschmidt echo}

 The LE is defined as
 the modulus square of the overlap $\mathcal{O}(t)$:
 \be \label{DLE}
   \mathcal{L}(t)=|\mathcal{O}(t)|^2
              =|\langle n|\psi(t)\rangle|^2,
 \ee
 which serves as the time-dependent fidelity and
 gives a measure of the instability of quantum evolution under small perturbation.
 Using the eigenstates of the final Hamiltonian, the initial eigenstate state $|n\rangle$ can be decomposed as
 $|n\rangle=\sum_k c_k|k^{f}\rangle$,where
 \be \label{ckc}
   c_k=\langle k^{f}|n\rangle
 \ee
 is the expansion coefficient and satisfies $\sum_k|c_k|^2=1$.
 Here, $|k^{f}\rangle$ denotes the $k$th eigenstate of the final Hamiltonian $\mathcal{H}_f$ with eigenenergy $E_k^f$.
 Then the LE in Eq.~(\ref{DLE}) can be rewritten in the following compact form:
 \be
   \mathcal{L}(t)=\left|\sum_k|c_k|^2 e^{-iE_k^f t}\right|^2.
 \ee

 Obviously, the time evolution of the LE is determined by the spectrum of the quenched Hamiltonian $\mathcal{H}_f$
 and its associated weight factors $c_k$.
 At $t=0$, according to the normalization condition of $c_k$, LE is equal to unity,
 while for $t>0$, the LE begins to decay.
 The value of the LE may approach zero under strong quenches.
 For finite-size systems, due to the finite-size effect, the LE will show a collapse and revival
 behavior after a sufficiently long time.
 However, the LE will eventually reach an asymptotic value $\mathcal{L}_\infty$ (equal to the long time average)
 for infinite-size systems \cite{campos}.

 \begin{figure*}
  \begin{minipage}[c]{0.6\textwidth}
    \includegraphics[width=\columnwidth]{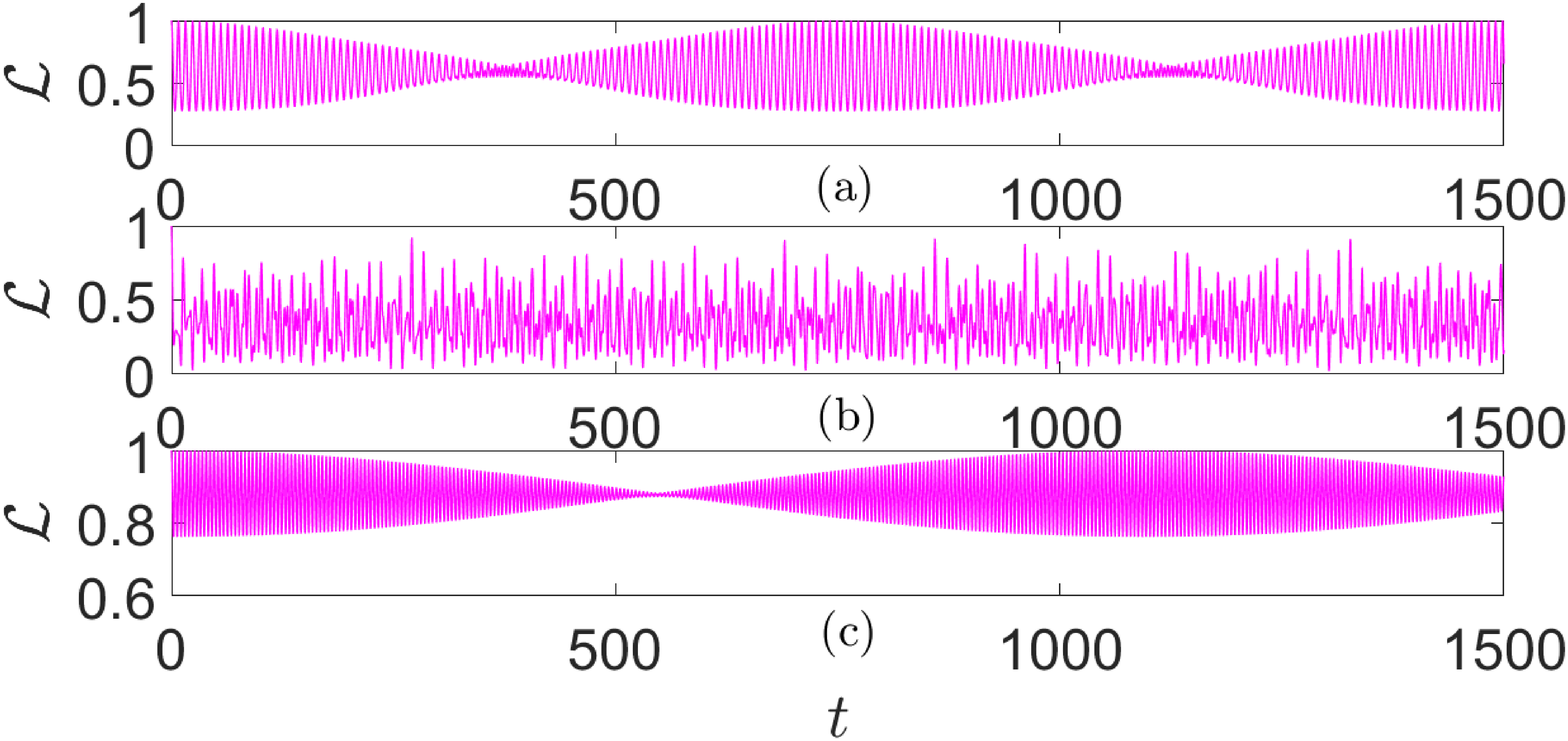}
  \end{minipage}%
  \hfill
  \begin{minipage}[c]{0.4\textwidth}
    \includegraphics[width=\columnwidth]{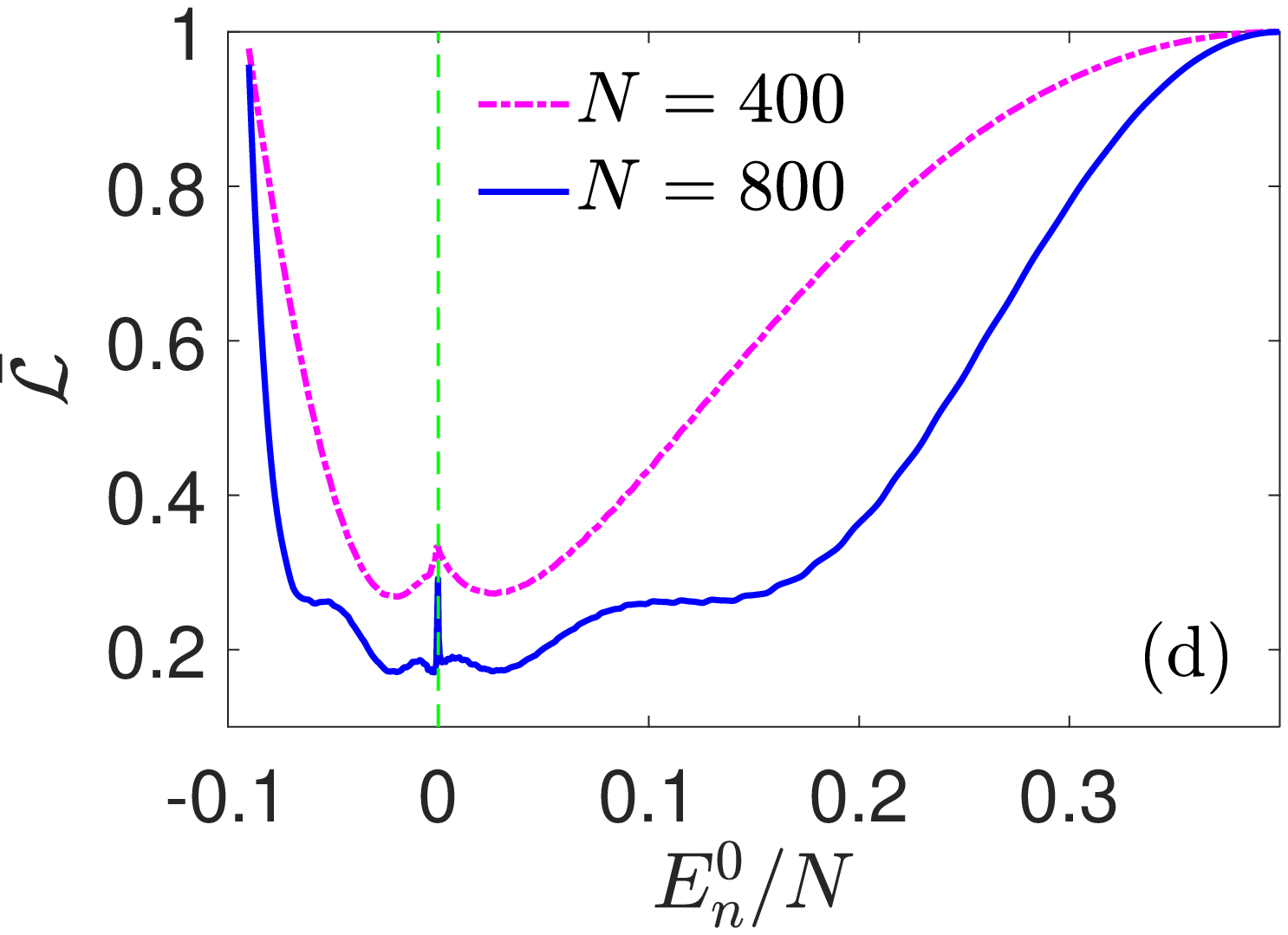}
  \end{minipage}
  \caption{(Color online)
   Time evolution of the LE for different initial states: (a) $|\mathrm{ini}=20\ra$
   with $E_{20}/N=-0.0468$, (b) $|\mathrm{ini}=80\ra$ with $E_{80}/N=3.8\times10^{-4}$, and (c) $|\mathrm{ini}=300\ra$ with $E_{300}/N=0.285$.
   Here, $N=400$, $\alpha=0.48$, and $\delta\alpha=0.01$.
   (d) $\bar{\mathcal{L}}$ as a function
   of the scaled eigenenergies of $\mathcal{H}_0$ for different $N$ with
   $\alpha=0.4$ and $\delta\alpha=0.01$.
   The vertical green dashed lines indicate the position of $E_c=0$.}
  \label{avLEeng}
 \end{figure*}

 In Fig.~\ref{TLE1}, we plot the LE for different controlling parameters with the initial state chosen to be the $77$th eigenstate and $N=400$.
 Here, according to the behavior of the second derivative of $E_{77}$ with respect to $\alpha$, one can find that, in this case,
 the critical point locates around $\alpha=0.48$.
 From Fig.~\ref{TLE1}, one can see that,
 in general, the LE shows a periodic oscillatory behavior for small quenches, while for the values of controlling
 parameter around $\alpha=0.48$ the LE exhibits more complicated behaviors and there is no periodicity.
 Moreover, for small values of $\alpha$, the LE periodically achieves orthogonality $(\mathcal{L}=0)$ [see Fig.~\ref{TLE1}(a)].
 Therefore, the underlying ESQPT has strong influences on the properties of the LE.

 In order to study the effects of ESQPT on the LE in a quantitative way,
 we numerically evaluate the probability distribution of the LE \cite{campos,keller}
 \be
    P(L)=\overline{\delta[L-\mathcal{L}(t)]}=\lim_{T\to\infty}\frac{1}{T}\int_0^T\delta[L-\mathcal{L}(t)]dt,
 \ee
 where $T$ is the total time of evolution.
 To calculate this distribution, the evolution time of the system should be chosen to be long
 in order to capture all the intricacies of the evolution.
 From the experiment point of view, one needs to prepare many copies of the initial states, evolve each of
 them independently, and measure the LE of individual states
 at different times $t_{i}$ \cite{campos}.
 Finally, from the measured data, the probability distribution of the LE can be constructed.
 In our simulation, we take $T=5000$, and we find that the results do not change for larger values of $T$.

 In Fig.~\ref{LED}, we plot the probability distribution of the LE for different controlling parameters.
 It can be seen that depending on the value of $\alpha$ the probability distributions
 of the LE exhibit quite different behaviors.
 The double-peaked distribution of the LE for small value of $\alpha$ implies the quasi-periodicity of the LE after a quench.
 However, when the value of $\alpha$ is large, the LE has a winged distribution which indicates the beating
 pattern behavior of the LE.
 Around the point $\alpha=0.48$, the probability distribution of the LE approximately obeys the Gaussian distribution.
 The time evolution of the LE exhibits a complex noisy behavior in this case.
 Here, we stress that increasing the system size $N$ with fixed $\delta\alpha$ will make the distribution of the LE
 approach an exponential one, which is similar to the Ising model shown in Ref.~\cite{campos}.
 However, for the large $N$ case, in order to investigate the different properties of the LE, the value of $\delta\alpha$
 also should be smaller.
 Therefore, qualitatively similar distribution shapes (see Fig.~\ref{LED}) can be obtained for any finite $N$.

 \begin{figure}
    \includegraphics[width=\columnwidth]{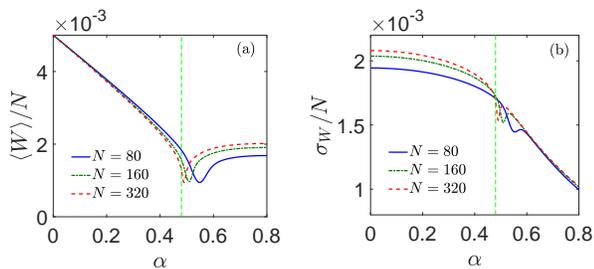}
  \caption{(Color online)
   (a) $\la W\ra/N$ as a function of $\alpha$ for different $N$.
   (b) $\sigma_W/N$ as a function of $\alpha$ for different $N$.
   In both figures the amplitude of the quench equals to $\delta\alpha=0.01$.
   The initial states are
   $|\mathrm{ini}=13\rangle$ for $N=80$, $|\mathrm{ini}=29\rangle$ for $N=160$,
   and $|\mathrm{ini}=61\rangle$ for $N=320$.
   The vertical green dashed lines indicate the critical point $\alpha_c=0.48$.}
  \label{averageW}
 \end{figure}

 To quantify the effects of ESQPT on the probability distribution of LE,
 we calculate the time-averaged value of the LE \cite{campos},
 \bey
   \bar{\mathcal{L}}=\int_{0}^1 LP(L)dL
                    =\lim_{T\to\infty}\frac{1}{T}\int_0^T\mathcal{L}(t)dt=\sum_k|c_k|^4.  \label{AVLE}
 \eey
 Obviously, the information of all excited states of $\mathcal{H}_f$ are incorporated in the averaged LE.
 In Fig.~\ref{average}, we show the averaged LE as a function of $\alpha$ for two different system sizes.
 It can be seen that in the vicinity of the critical point the averaged LE has a cusp, thus signaling the ESQPT.
 Therefore, the signatures of ESQPT can be revealed by the statistical properties of the LE.
 It is worth pointing out that for fixed $\delta\alpha$, the decaying of the LE will be enhanced by
 increasing the system size $N$ (cf.~Fig.~\ref{average}).

 Here, we stress that although the point $\alpha=0.48$ is obtained from $77$th eigenstate with $N=400$,
 the critical point of ESQPT is fixed for different system size.
 We find that for any $N$, the number of the
 eigenstate which shows an ESQPT around $\alpha=0.48$ is given by $|n\rangle=|N/5-3\rangle$.

 The ESQPT can also be characterized by the singularity in the density of states at the critical energy with fixed
 controlling parameters.
 Therefore, we need to study if the critical energy of the ESQPT can be determined by the statistics of the LE.
 In Figs.~\ref{avLEeng}(a)-\ref{avLEeng}(c), we plot the time evolution of LE for different
 initial states with fixed $\alpha$ and $\delta\alpha$.
 For $E>E_c=0$ and $E<E_c$, we can see the regular oscillatory behaviors of the LE.
 By contrast, for $E\approx E_c=0$, we see that the time evolution of the LE is irregular,
 and the value of LE cannot reach unity dynamically.
 Similar to the different controlling parameter cases, the probability distribution of the LE will exhibit different
 behaviors for different initial energies.
 The critical energy of ESQPT therefore can be probed through the statistical properties of the LE.
 Indeed, the averaged value of the LE shows a cusp near the critical energy $E_c=0$ [see Fig.~\ref{avLEeng}(d)].
 The cusp becomes sharper as the size of system increases
 and its location approaches $E_c=0$ as $N\to\infty$.
 Note that there are many of similarities between the behaviors of the LE
 at $\alpha\approx\alpha_c$ and at $E_n^0\approx E_c$.

 \subsection{Quantum work distribution}

 In this subsection, we study the ESQPT through the work $W$ done on the system during a sudden quench process.
 We first give a brief review of the work probability distribution $P(W)$, point out the relation between $P(W)$ and $\mathcal{O}$
 in Eq.~(\ref{TOL}), and illustrate the key findings with our numerical results.

 For the sudden quench process,
 the Hamiltonian of the system before and after the quench can be written in the following form:
 $\mathcal{H}_0=\sum_n E_n^0|n\rangle\langle n|$, $\mathcal{H}_f=\sum_{k} E_{k}^f|k^{f}\rangle\langle k^{f}|$.
 In order to write down a simple expressions of the work distribution and the $l$th moment of the work distribution,
 we assumed that there is no degeneracy in the eigenvalues.
 However, in our numerical simulations, the degeneracy in eigenvalues has been considered.
 The work distribution of this process can be written as \cite{talkner,fusco}
 \be \label{PDW}
   P(W):=\sum_{n,k}p^0_n p(k^{f}|n)\delta[W-(E_{k}^f-E_n^0)].
 \ee
 Here, $p^0_n=\mathrm{Tr}[P_n^0\rho_0]$ denotes the probability with which the energy value $E_n^0$ is observed in the initial
 energy measurement,
 and $P_n^0=|n\rangle\langle n|$ is the projection operator.
 The probability of obtaining the eigenvalue $E_{k}^f$ at the final moment of time, conditioned on the observation
 of $E_n^0$ at the initial time, is given by
 $p(k^{f}|n)=|\langle k^{f}|\hat{U}|n\rangle|^2$,
 where $\hat{U}$ is the unitary evolution operator.

 \begin{figure}
    \includegraphics[width=\columnwidth]{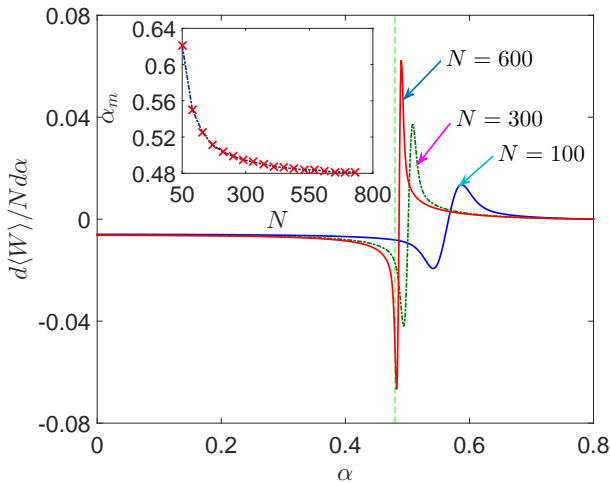}
    \caption{(Color on line)
    The derivative of the scaled averaged work as a function of $\alpha$ for different $N$
    with $\delta\alpha=0.01$.
    The initial states are:
    $|\mathrm{ini}=17\rangle$ for $N=100$, $|\mathrm{ini}=57\rangle$ for $N=300$,
    and $|\mathrm{ini}=117\rangle$ for $N=600$.
    The vertical green dashed line indicates the critical point $\alpha_c=0.48$.
    Inset: exact position $\alpha_m$ (corresponding to the minimum in $d[\langle W\rangle/N]/d\alpha$) as a
    function of $N$.
    The initial state for each $N$ is $|\mathrm{ini}=N/5-3\ra$.}
    \label{scale}
 \end{figure}%

 From the work distribution (\ref{PDW}), the characteristic function of work, defined
 as the Fourier transformation of $P(W)$ \cite{fusco,silva,talkner}, can be expressed as
 \begin{align}
    \chi(t)&=\int dW e^{itW}P(W), \nonumber  \\
           &=\sum_{n,k}\exp{[i(E_{k}^f-E_n^0)t]}|\langle k^{(f)}|\hat{U}|n\rangle|^2p_n^0.
 \end{align}
 Because $\hat{U}$ equals to the identity operator in the sudden quench process,
 $\chi(t)$ then is reduced to the complex conjugate of the time-dependent overlap $\mathcal{O}(t)$ with
 the average taken on the initial state $\rho_0$ \cite{silva}.

 The $l$th moment of the work distribution in the sudden quench process can be obtained through the $l$th derivative of $\chi(t)$ with
 respect to $t$ at $t=0$, and the result is \cite{fusco}
 \be \label{avgW}
   \langle W^l\rangle=\mathrm{Tr}\left[\sum_{m=0}^l(-1)^m\binom{l}{m}\mathcal{H}_f^{(l-m)}\mathcal{H}_0^l\rho'_0\right],
 \ee
 where $\forall{l}\in\mathbb{N}$ and $\rho'_0=\sum_n P_n^0\rho_0 P_n^0$ is the initial projected state.
 For the sudden quench process starting from the $n$th eigenstate of the system, Eq.~(\ref{avgW}) reads
 \be
   \langle W^l\rangle=\sum_{k}|c_k|^2[E_{k}^f-E_n^0]^l.
 \ee

 Let us study the effects of ESQPT on the statistics of the work $W$
 during a sudden quench process.
 In Fig.~\ref{averageW}(a), we plot the averaged work as a function of $\alpha$ for different $N$.
 For every $N$ we choose the eigenstate that exhibits an ESQPT at about $\alpha_c=0.48$.
 Clearly, it can be seen that the averaged work shows a cusp around $\alpha_c$.
 Moreover, as the size $N$ of the system increases, the cusp of the averaged work becomes sharper.
 The exact position of the cusp changes with the size $N$ of the system and
 approaches $\alpha_c$ as $N\to\infty$ \cite{barbar}.
 The standard deviation of work distribution $\sigma_W=\sqrt{\langle W^2\rangle-\langle W\rangle^2}$ as
 a function of $\alpha$ has been plotted in Fig.~\ref{averageW}(b).
 Similar to the averaged work, around $\alpha_c$ the standard deviation of work distribution also exhibits a cusp,
 its location approaches $\alpha_c$ as $N$ increases.
 Both the averaged work and the standard deviation of the work distribution exhibit a sharp drop at the critical point of ESQPT.

 \begin{figure}
    \includegraphics[width=\columnwidth]{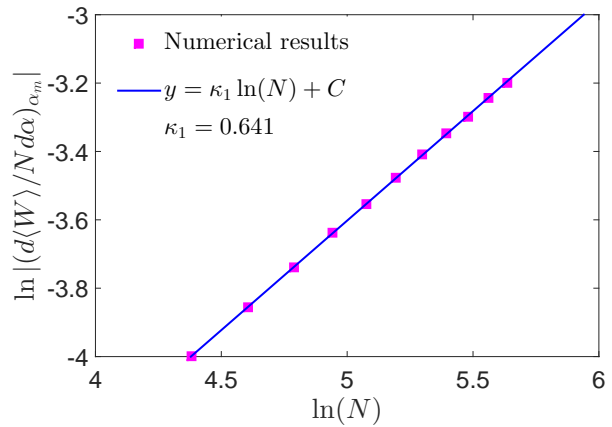}
    \caption{(Color on line)
    The absolute value of the minimum $d[\langle W\rangle/N]/d\alpha$ (in logarithmic scale) as a function of $N$ (in logarithmic scale).
    The amplitude of the quench is $\delta\alpha=0.01$, and the initial state for each $N$ is $|\mathrm{ini}=N/5-3\rangle$.}
    \label{scaledW}
 \end{figure}%

 To study the scaling behaviors of the averaged work, we plot the derivative of the averaged
 work with respect to $\alpha$ as a function of $\alpha$ for different $N$ in Fig.~\ref{scale}.
 One can see that the derivative of the averaged work has a minimum around $\alpha_c=0.48$.
 The amplitude of the minimum is significantly enhanced by increasing the system size $N$.
 Moreover, the location $\alpha_m$ of the minimum in $d[\langle W\rangle/N]/d\alpha$, which can be regarded as the precursor
 of the critical point $\alpha_c$, moves towards the critical point when $N$ increases
 and approaches $\alpha_c=0.48$ as $N\to\infty$ (see the inset of Fig.~\ref{scale}).

 \begin{figure}
    \includegraphics[width=\columnwidth]{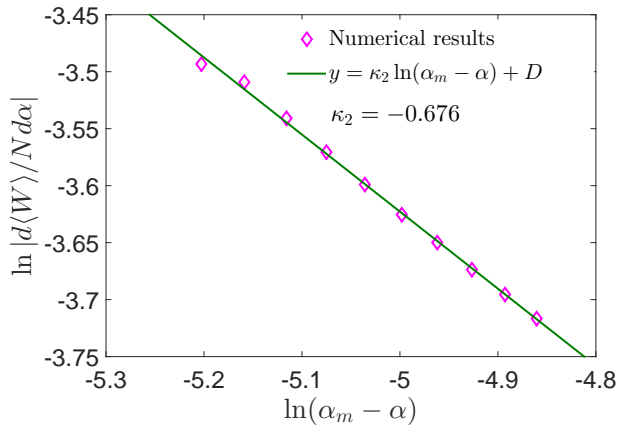}
    \caption{(Color on line)
    The absolute values of $d[\langle W\rangle/N]/d\alpha$ (in logarithmic scale) as a function of $\alpha_m-\alpha$ (in logarithmic scale).
    The amplitude of the quench is $\delta\alpha=0.01$, the system size is $N=800$, and the initial state is $|\mathrm{ini}=157\rangle$.}
    \label{chgedW}
 \end{figure}%

 The absolute value of the minimum $d[\langle W\rangle/N]/d\alpha$ as a
 function of $N$ is plotted in Fig.~\ref{scaledW}.
 Clearly, one can see that as the system size $N$ increases, the value of
 the minimum of the derivative of the averaged work diverges as
 \be
   \ln\left|\left(\frac{d\langle W\rangle}{N d\alpha}\right)_{\alpha_m}\right|=\kappa_1\ln N+C,
 \ee
 where $C$ is a constant and $\kappa_1=0.641$.
 To get more information about the ESQPT from the averaged work, we plot the
 behavior of $\ln[|d(\langle W\rangle/N)/d\alpha|]$ in the vicinity of $\alpha_m$ for a large $N$ case in Fig.~\ref{chgedW}.
 From the figure, we found that in the neighborhood of $\alpha_m$, $\ln[|d(\langle W\rangle/N)/d\alpha|]$ has the
 following asymptotic expression:
 \be
   \ln\left|\frac{d\langle W\rangle}{Nd\alpha}\right|=\kappa_2\ln(\alpha_m-\alpha)+D,
 \ee
 where $\kappa_2=-0.676$ and $D$ is a constant.
 According to the finite-size scaling theory of the phase transition \cite{barbar}, the value of the critical exponent $\nu_e$
 is given by $\nu_e=|\kappa_2/\kappa_1|\approx1$ \cite{foot}.
 Here, we should point out that although the values of $\kappa_1$ and $\kappa_2$ are dependent on controlling parameters,
 $\nu_e$ is a constant for a given system.

 \begin{figure}
    \includegraphics[width=\columnwidth]{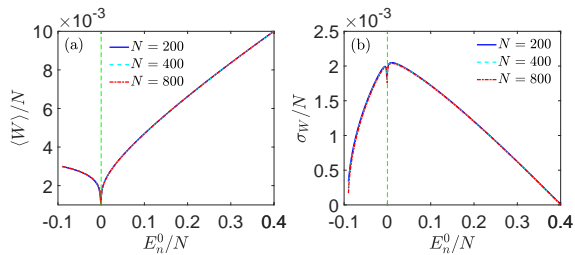}
  \caption{(Color online)
   (a) Rescaled averaged work as a function of $E^0_n/N$ for different $N$.
   (b) Rescaled variance of work as a function of $E_n^0/N$ for different $N$.
   In both figures, the controlling parameter is $\alpha=0.4$ and $\delta\alpha=0.01$.
   The vertical green dashed lines indicate the critical energy $E_c=0$.}
  \label{averageWE}
 \end{figure}

 We further investigate the averaged work and the standard deviation of work distribution for different initial states.
 In Fig.~\ref{averageWE}, we plot the averaged work and the standard deviation of work distribution as a function of $E_n^0$.
 Obviously, in the neighbourhood of the critical energy, both the averaged work and the standard deviation
 of work show a cusplike shape, and become sharper as $N$ increase.
 The locations of the cusp in $\langle W\rangle$ and $\sigma_W$ change as the size of the system grows
 and approach an asymptotic value $E_c=0$ as $N\to\infty$.
 Finally, we remark that qualitatively very similar results can be obtained for any other higher moment of the work distribution.
 Therefore, the signature of the ESQPT can be captured by the work during a sudden quench process.

 It is worth pointing out that the right panels of Figs.~\ref{averageW} and \ref{averageWE} look very similar.
 Actually, Fig.~\ref{averageW} displays the influence of ESQPT on the averaged work and the standard deviation of work for specific excited states,
 with the different controlling parameter. Figure~\ref{averageWE} exhibits the influence of ESQPT on the averaged work and the standard deviation of work
 for every excited states with fixed controlling parameter.
 The results shown in these two figures correspond to two different definitions of the ESQPT.

 \section{Spectral function and Localization} \label{sfl}

\begin{figure*}
  \begin{minipage}[t]{0.33\linewidth}
    \centering
    \includegraphics[width=\columnwidth]{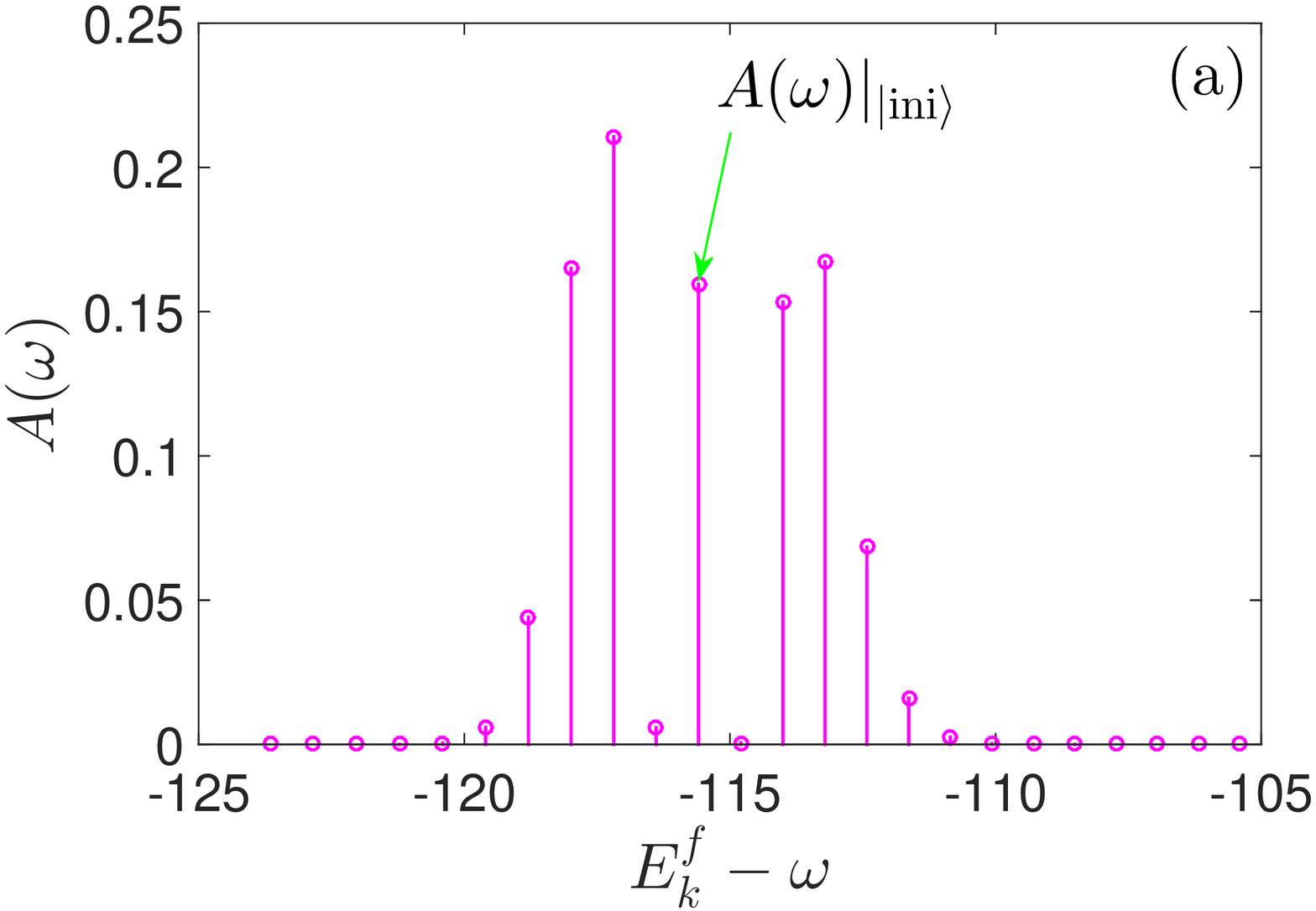}
  \end{minipage}%
  \hfill
  \begin{minipage}[t]{0.33\linewidth}
    \centering
    \includegraphics[width=\columnwidth]{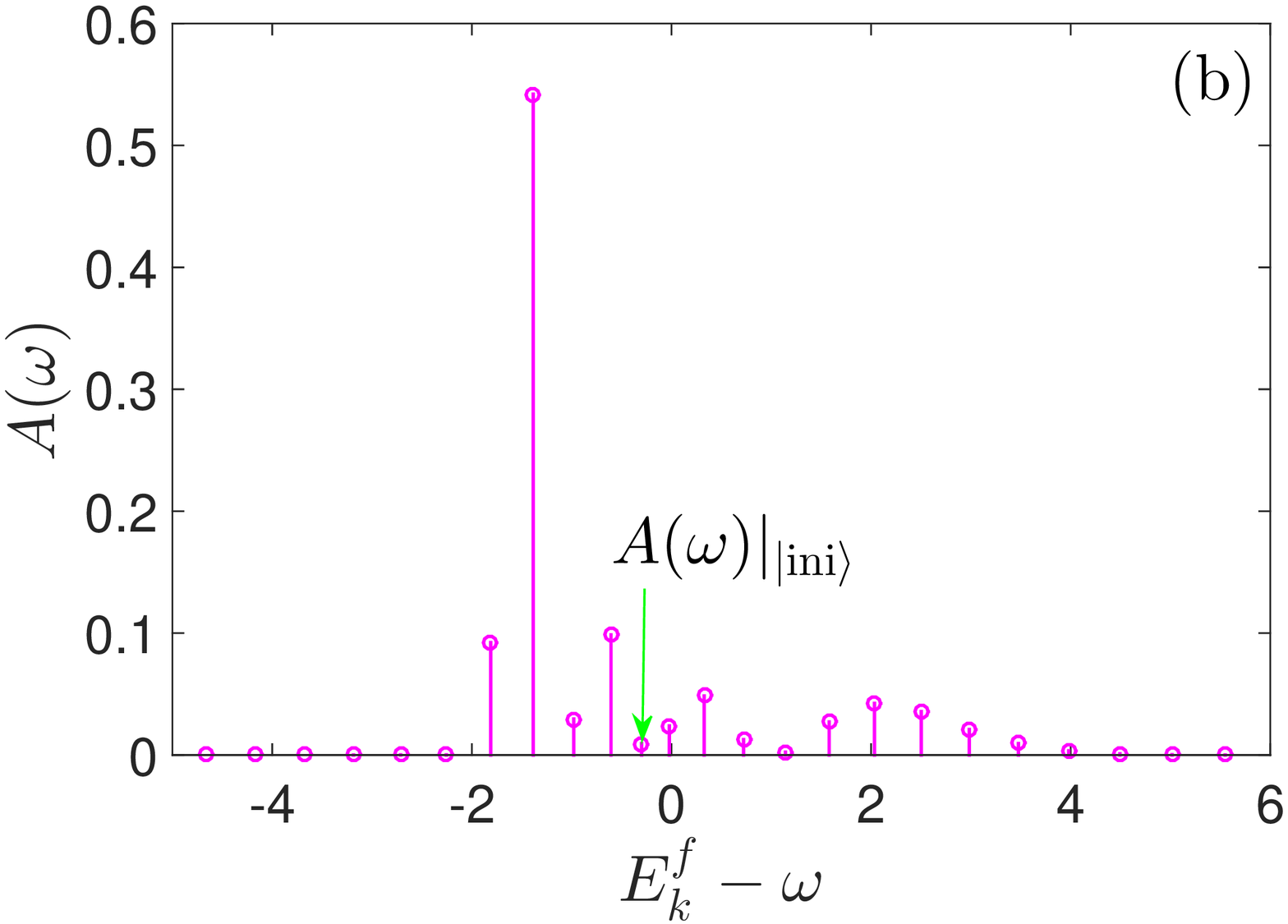}
  \end{minipage}%
  \hfill
  \begin{minipage}[t]{0.33\linewidth}
    \centering
    \includegraphics[width=\columnwidth]{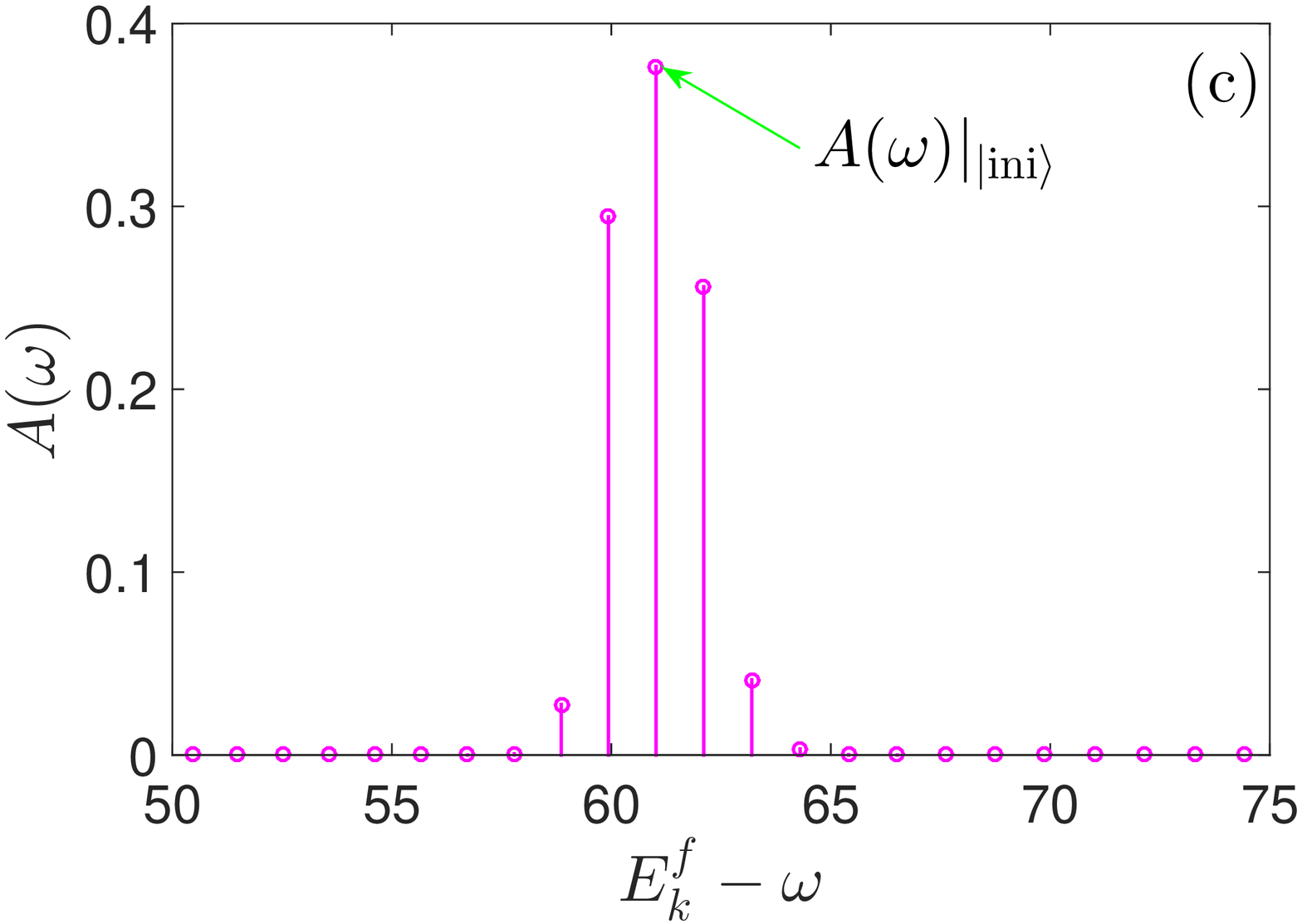}
  \end{minipage}%
  \caption{(Color online)
   Spectral function for different controlling parameters: $\alpha=0.1$ (a), $\alpha=0.48$ (b), and $\alpha=0.8$ (c), with
   $N=1000$ and $\delta\alpha=0.01$. The initial state is $|\mathrm{ini}=197\rangle$.
   The green arrows indicate the value of $|A(\omega)|$ corresponding to the initial state.}
  \label{SFF}
 \end{figure*}

 To get a better understanding of the nonequilibrium dynamics of the system, in this section we do a spectral analysis
 of $\mathcal{O}(t)$.
 For the sudden quench process, the spectral function, which is defined as the real part of Fourier transformation
 of $\mathcal{O}(t)$ \cite{campos,keller,campbell}, can be written as
 \bey
    A(\omega)&=&\mathfrak{R}\left[\int_{-\infty}^{\infty}dt e^{i\omega t}\mathcal{O}(t)\right], \nonumber \\
             &=&\sum_{k}|c_k|^2\delta[\omega-(E_{k}^f-E_n^0)]. \label{SF}
 \eey
 It describes the fundamental excitations that govern the subsequent quantum dynamics.
 Obviously, it is related to the work distribution probability in Eq.~(\ref{PDW}).
 We plot spectral functions for several controlling parameters: $\alpha<\alpha_c, \alpha=\alpha_c$, and $\alpha>\alpha_c$, in Fig.~\ref{SFF}.
 We stress that even though our results are obtained for $N=1000$, qualitatively similar results can be obtained for much larger systems.

 In Fig.~\ref{SFF}(a), we take $\alpha=0.1$. It can be seen that in this case the spectral function spreads over several energy levels.
 The highest peak does not correspond to the initial state.
 This implies that the dynamics of the system is dominated by several eigenstates, which results in dynamical orthogonality.
 The large values of the averaged work and the standard deviation of work can also be explained by this property of the spectral function.
 The peak corresponding to the initial state has approximately the same amplitude as peaks corresponding to other excited states.
 This in turn allows us to understand
 the periodic behavior of the LE.

 In Fig.~\ref{SFF}(c), $\alpha=0.8$. One can also see that several excited states contribute to the dynamics of the system.
 However, in this case, the highest peak corresponds to the initial state.
 These properties of the spectral function explain why one cannot witness the dynamic orthogonality in the behavior of LE,
 even though the behavior of the LE is periodic.
 Compared with Fig.~\ref{SFF}(a), the width of the spectral function is narrower.
 Therefore, both of the averaged work and the standard deviation
 of work distribution have a smaller value in this case.

 In Fig.~\ref{SFF}(b), we show the critical case with $\alpha=\alpha_c=0.48$.
 The localization of the spectral function implies that the averaged work and the standard deviation of work
 distribution sharply decay at the critical point.
 The small contributions of the initial state and some excited states to $A(\omega)$
 explain why the time evolution of the LE is complex and why there is no periodicity at the critical point.
 Finally, it is worth pointing out that depending on the parity of the system, the Hilbert space can be
 divided into two subspaces with even and odd parities, respectively.
 Therefore, only the states which have the same parity as the initial state play a role in the dynamics.

 The features of the spectral function shown in Fig.~\ref{SFF} strongly suggest that the initial state of the system
 is localized at the critical point of ESQPT.
 To verify this conjecture, we evaluate the inverse participation ratio (PR) \cite{santos},
 which measures the localization of a state in a chosen basis.
 For the case that we study here, PR is defined as
 \be
   \mathrm{PR}_{\mathcal{H}_f}^{(\mathrm{ini})}=\frac{1}{\sum_k|c_k|^4}, \label{PRE}
 \ee
 where $c_k$ is given by Eq.~(\ref{ckc}).
 We should point out that depending on the parity of the initial state, the sum in Eq.~(\ref{PRE}) involves either even or odd values of $k$.

 A localized state will lead to a small value of PR, while a delocalized state gives a large value of PR.
 Comparing with Eq.~(\ref{AVLE}), we find an interesting relation between the averaged LE and PR, i.e.,
 $\mathrm{PR}_{\mathcal{H}_f}^{(\mathrm{ini})}=1/\bar{\mathcal{L}}$.
 Therefore, at the critical point, the cusplike peak in the averaged LE will induce a cusplike dip in PR.
 This means that PR will decrease sharply at the critical point.

 \begin{figure}
    \includegraphics[width=\columnwidth]{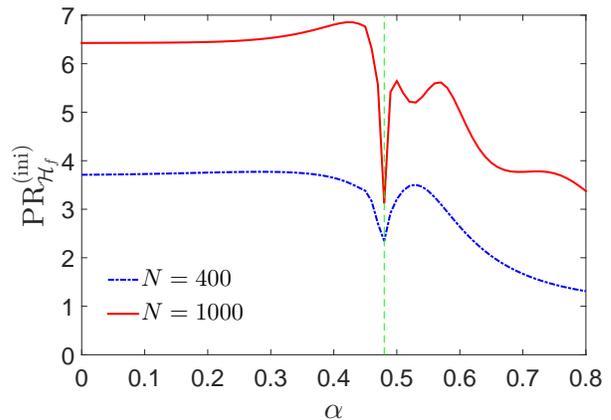}
    \caption{(Color on line)
    Participation ratio (PR) as a function of $\alpha$ for different sizes of system with $\delta\alpha=0.01$.
    The initial states are: $|\mathrm{ini}=77\rangle$ for $N=400$,
    $|\mathrm{ini}=197\rangle$ for $N=1000$.
    The vertical green dashed line indicates the critical point $\alpha_c=0.48$.}
    \label{PRa}
 \end{figure}%

 In Fig.~\ref{PRa}, we show $\mathrm{PR}_{\mathcal{H}_f}^{(\mathrm{ini})}$ as a function of $\alpha$ for two different $N$
 with different initial states that have an ESQPT at $\alpha_c=0.48$ (see caption for details).
 It is obvious that a pronounced dip become noticeable in the neighbourhood of the critical point.
 Moreover, as the size of system increases, the dip becomes more pronounced and the location of the
 dip moves toward the critical point.
 Therefore, the inverse PR can be used as a useful tool to detect ESQPT.

\begin{figure}
    \includegraphics[width=\columnwidth]{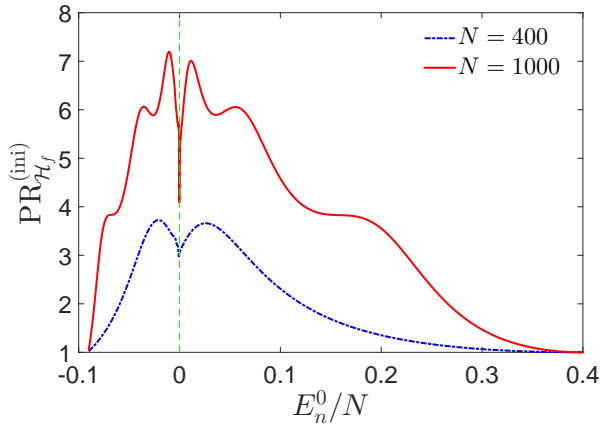}
    \caption{(Color on line)
    Participation ratio (PR) of all the initial states of the even parity sector;
    $E_n^0/N$ is the scaled eigenenergy of the $n$th initial state.
    The parameters are: $\alpha=0.4$ and $\delta\alpha=0.01$.
    The vertical green dashed line indicates the critical energy $E_c=0$.}
    \label{PREg}
 \end{figure}

 We also plot $\mathrm{PR}_{\mathcal{H}_f}^{(\mathrm{ini})}$ for all eigenstates of $\mathcal{H}_0$
 as a function of scaled initial eigenenergy $E_n^0/N$ in Fig.~\ref{PREg}.
 Here, we fix $\alpha=0.48$ and $\delta\alpha=0.01$.
 We can see clearly at the edges of the spectrum that the eigenstates are localized with small values of PR.
 Particularly, for the initial states with energies close to the critical energy $E_c=0$, the value of PR has a dip,
 which becomes more pronounced as the size of system $N$ increases.
 From these results we can confirm that at the critical point of the ESQPT the initial state of
 the system becomes a localized state and PR serves as a good indicator of ESQPTs \cite{santos}.

 \section{Conclusions} \label{sum}

 In conclusion, we have studied in detail the effects of an ESQPT on the nonequilibrium dynamics of a quantum
 many-body system, i.e., LMG model, by quenching its controlling parameter.
 Unlike the studies in Refs.~\cite{relano,perez,santos}, in our study the effects of
 the ESQPT are analyzed through the statistics of the LE and quantum work.
 We set the initial state to be the $n$th ($n>1$) eigenstate of the system and fix the amplitude of the quench.
 We have shown that the probability distributions of the LE and work exhibit
 distinct behaviors when the controlling parameter locates at and away from the critical point.
 For different initial states, the same phenomenon can be observed.

 The probability distribution of the LE shows a double-peaked or winged shape when the
 controlling parameter is far away from the critical value.
 While at the critical point, the distribution of the LE is approximately given by a Gaussian distribution.
 To quantify the effects of the ESQPT on the statistics of the LE, we studied the averaged LE.
 The cusplike shape of the averaged LE
 with a singularity localized at the critical point $\alpha_c/E_c$ has been found.
 By analyzing the quantum work distribution, we have found that both of the
 averaged work and the standard deviation of work can be used to characterize the ESQPT.
 The scaling behavior of the averaged work around the critical point of ESQPT has been studied.
 We obtained the critical exponent $\nu_e$, which characterizes the divergence of the correlation length \cite{foot} near the critical point.

 To fully understand the influence of ESQPT on the nonequilibrium dynamics in the LMG model, the spectral function was analyzed.
 We have found that the inverse PR shows a dip at the critical point.
 Hence, at the critical point, the initial state becomes a localized state.
 Our results highlight the relation between ESQPT and the nonequilibrium quantum dynamics,
 induced by quenching the controlling parameters of a quantum system.
 Unveiling how the nonequilibrium dynamics is influenced by ESQPTs may provide more understanding about the ESQPT.

 \acknowledgements

 H.~T.~Q. gratefully acknowledges support from the National Science Foundation of China
 under Grants No.~11375012 and No.~11534002 and the Recruitment
 Program of Global Youth Experts of China.


\begin{thebibliography}{99}

\bibitem{sachdev} S.~Sachdev, {\it Quantum Phase Transitions} (Cambridge University Press, Cambridge, 1999).

\bibitem{vojta} M.~Vojta, Rep.~Prog.~Phys. {\bf 66}, 2069 (2003).

\bibitem{dutta} A.~Dutta, G.~Aeppli, B.~K.~Chakrabarti, U.~Divakaran, T.~F.~Rosenbaum, and D.~Sen, {\it Quantum Phase Transition in
                Transverse Field Spin Models: From Statistical Physics to Quantum Information} (Cambridge University Press, Cambridge, 2015).

\bibitem{suzuki} M.~Suzuki, Prog.~Theor.~Phys. {\bf 46}, 1337 (1971).

\bibitem{amico} L.~Amico, R.~Fazio, A.~Osterloh, and V.~Vedral, Rev.~Mod.~Phys. {\bf 80}, 517 (2008).

\bibitem{zanardi} P.~Zanardi and N.~Paunkovic, Phys.~Rev.~E {\bf 74}, 031123 (2006).

\bibitem{emary} C.~Emary and T.~Brandes, Phys.~Rev.~Lett. {\bf 90}, 044101 (2003).

\bibitem{dutta2} V.~Mukherjee, S.~Sharma, and A.~Dutta, Phys.~Rev.~B {\bf 86}, 020301 (2012).

\bibitem{haitao} H.~T.~Quan, Z.~Song, X.~F.~Liu, P.~Zanardi, and C.~P.~Sun, Phys.~Rev.~Lett. {\bf 96}, 140604 (2006).

\bibitem{wyr} Wang Yao, Ren-Bao Liu, and L.~J.~Sham, Phys.~Rev.~B {\bf 74}, 195301 (2006).

\bibitem{rossini} D.~Rossini, T.~Calarco, V.~Giovannetti, S.~Montangero, and R.~Fazio, Phys.~Rev.~A {\bf 75}, 032333 (2007).

\bibitem{paz} F.~M.~Cucchietti, S.~Fernandez-Vidal, and J.~P.~Paz, Phys.~Rev.~A {\bf 75}, 032337 (2007).

\bibitem{zurek} W.~H.~Zurek, U.~Dorner, and P.~Zoller, Phys.~Rev.~Lett. {\bf 95}, 105701 (2005).

\bibitem{polk} A.~Polkovnikov, Phys.~Rev.~B {\bf 72}, 161201(R) (2005).

\bibitem{polkovnikov} A.~Polkovnikov, K.~Sengupta, A.~Silva, and M.~Vengalattore, Rev.~Mod.~Phys. {\bf 83}, 863 (2011).

\bibitem{agsilva} A.~Gambassi, A.~Silva, Phys.~Rev.~Lett. {\bf 109}, 250602 (2012).

\bibitem{dgcm} R.~Dorner, J.~Goold, C.~Cormick, M.~Paternostro, and V.~Vedral, Phys.~Rev.~Lett. {\bf 109}, 160601 (2012).

\bibitem{suter} Jingfu Zhang, Xinhua Peng, N.~Rajendran, and D.~Suter, Phys.~Rev.~Lett. {\bf 100}, 100501 (2008).

\bibitem{ditty} Jingfu Zhang, F.~M.~Cucchietti, C.~M.~Chandrashekar, M.~Laforest, C.~A.~Ryan,
           M.~Ditty, A.~Hubbard, J.~K.~Gamble, and R.~Laflamme, Phys.~Rev.~A {\bf 79}, 012305 (2009).

\bibitem{greiner} M.~Greiner, O.~Mandel, T.~Esslinger, T.~W.~Hansch, and I.~Bloch, Nature (London) {\bf 415}, 39 (2002).

\bibitem{bloch} I.~Bloch, J.~Dalibard, and W.~Zwerger, Rev.~Mod.~Phys. {\bf 80}, 885 (2008).

\bibitem{baumann} K.~Baumann, C.~Guerlin, F.~Brennecke, and T.~Esslinger, Nature (London) {\bf 464}, 1301 (2010).

\bibitem{macek} P.~Cejnar, M.~Macek, S.~Heinze, J.~Jolie, and J.~Dobes, J.~Phys.~A {\bf 39}, L515 (2006).


\bibitem{caprio} M.~A.~Caprio, P.~Cejnar, and F.~Iachello, Ann.~Phys. (NY) {\bf 323}, 1106 (2008).

\bibitem{psmpc} P.~Stransky, M.~Macek, and P.~Cejnar, Ann.~Phys. (NY) {\bf 345}, 73 (2014).

\bibitem{ribeiro1} P.~Ribeiro, J.~Vidal, and R.~Mosseri, Phys.~Rev.~Lett. {\bf 99}, 050402 (2007).

\bibitem{ribeiro2} P.~Ribeiro, J.~Vidal, and R.~Mosseri, Phys.~Rev.~E {\bf 78}, 021106 (2008).

\bibitem{zigang} Z.~G.~Yuan, P.~Zhang, S.~S.~Li, J.~Jing, and L.~B.~Kong, Phys.~Rev.~A {\bf 85}, 044102 (2012).

%
%


\bibitem{relano} A.~Relano, J.~M.~Arias, J.~Dukelsky, J.~E.~Garc\'{i}-Ramos, and P.~P\'{e}rez-Fernandez, Phys.~Rev.~A {\bf 78}, 060102(R) (2008).

\bibitem{perez} P.~P\'{e}rez-Fernandez, A.~Relano, J.~M.~Arias, J.~Dukelsky, and J.~E.~Garc\'{i}-Ramos, Phys.~Rev.~A {\bf 80}, 032111 (2009).

\bibitem{santos} L.~F.~Santos, M.~Tavora, and F.~P.~Bernal, Phys.~Rev.~A {\bf 94}, 012113 (2016).

\bibitem{cejnar} P.~Cejnar and J.~Jolie, Prog.~Part.~Nucl.~Phys. {\bf 62}, 210 (2009).


%
%

\bibitem{pcrf} P.~Cejnar, J.~Jolie, and R.~F.~Casten, Rev.~Mod.~Phys. {\bf 82}, 2155 (2010).  

%

\bibitem{brandes} T.~Brandes, Phys.~Rev.~E {\bf 88}, 032133 (2013).

\bibitem{mabm} M.~A.~Bastarrachea-Magnani, S.~Lerma-Hernandes, and J.~G.~Hirsch, Phys.~Rev.~A {\bf 89}, 032101 (2014).



\bibitem{puebla2} R.~Puebla, Myung-Joong Hwang, and M.~B.~Plenio, Phys.~Rev.~A {\bf 94}, 023835 (2016).

\bibitem{bastidas} V.~M.~Bastidas, P.~Perez-Bernal, M.~Vogl, and T.~Brandes, Phys.~Rev.~Lett. {\bf 112}, 140408 (2014).

\bibitem{dietz} B.~Dietz, F.~Iachello, M.~Miski-Oglu, N.~Pietralla, A.~Richter, L.~von Smekal, and J.~Wambach, Phys.~Rev.~B {\bf 88}, 104101 (2013).

\bibitem{bpw} B.~P.~Winnewisser, M.Winnewisser, I.~R.~Medvedev, M.~Behnke, F.~C.~De Lucia, S.~C.~Ross, and J.~Koput, Phys.~Rev.~Lett. {\bf 95}, 243002 (2005).

\bibitem{zobov} N.~F.~Zobov, S.~V.~Shirin, O.~L.~Polyansky, J.~Tennyson, P.~F.~Coheur, P.~F.~Bernath, M.~Carleer, and R.~Colin, Chem.~Phys.~Lett. {\bf 414}, 193 (2005).

\bibitem{larese} D.~Larese and F.~Iachello, J.~Mol.~Struct. {\bf 1006}, 611 (2011).

\bibitem{larese2} D.~Larese, F.~Perez-Bernal, and F.~Iachello, J.~Mol.~Struct. {\bf 1051}, 310 (2013).

\bibitem{larese3} D.~Larese, M.~A.~Caprio, F.~Perez-Bernal, and F.~Iachello, J.~Chem.~Phys. {\bf 140}, 014304 (2014).

\bibitem{dietz2} F.~Iachello, B.~Dietz, M.~Miski-Oglu,and A.~Richter, Phys.~Rev.~B {\bf 91}, 214307 (2015).  

\bibitem{ardp} A.~Relano, J.~Dukelsky, P.~P-Fernandez, and J.~M.~Arias, Phys.~Rev.~E {\bf 90}, 042139 (2014).

\bibitem{puebla} R.~Puebla, A.~Relano, and J.~Retamosa, Phys.~Rev.~A {\bf 87}, 023819 (2013).



\bibitem{gevm} G.~Engelhardt, V.~M.~Bastidas, W.~Kopylov, and T.~Brandes, Phys.~Rev.~A {\bf 91}, 013631 (2015).

\bibitem{wktb} W.~Kopylov and T.~Brandes, New.~J.~Phys. {\bf 17}, 103031 (2015).


\bibitem{lobez} C.~M.~Lobez and A.~Relano, Phys.~Rev.~E {\bf 94}, 012140 (2016).





\bibitem{Lipkin} H.~J.~Lipkin, N.~Meshkov, and A.~J.~Glick, Nucl.~Phys. {\bf 62}, 188 (1965).


\bibitem{dusuel2} S.~Dusuel and J.~Vidal, Phys.~Rev.~B {\bf 71}, 224420 (2005).


\bibitem{botet1} R.~Botet and R.~Jullien, Phys.~Rev.~B {\bf 28}, 3955 (1983).   

\bibitem{htzc} H.~T.~Quan, Z.~D.~Wang, and C.~P.~Sun, Phys.~Rev.~A {\bf 76}, 012104 (2007).

\bibitem{qianwang} Q.~Wang, P.~Wang, Y.~B.~Yang, and W.~G.~Wang, Phys.~Rev.~A {\bf 91}, 042102 (2015).





\bibitem{jalabert} R.~A.~Jalabert and H.~M.~Pastawski, Phys.~Rev.~Lett. {\bf 86}, 1490 (2001).

\bibitem{gorin} T.~Gorin, T.~Prosenb, T.~H.~Seligmanc, M.~Znidaric, Phys.~Rep {\bf 435}, 33 (2006).

\bibitem{fusco} L.~Fusco {\it et al}, Phys.~Rev.~X {\bf 4}, 031029 (2014).

\bibitem{dasone} S.~Bhattacharyya, S.~Dasgupta, and A.~Das, Sci.~Rep. {\bf 5}, 16490 (2015).

\bibitem{dastwo} S.~Roy, R.~Moessner, and A.~Das, Phys.~Rev.~B {\bf 95}, 041105(R) (2017).

\bibitem{halimeh} I.~Homrighausen, N.~O.~Abeling, V.~Z.~Stauber, and J.~Halimeh, arXiv:1703.09195v2.

\bibitem{jafari} R.~Jafari and H.~Johannesson, Phys.~Rev.~Lett. {\bf 118}, 015701 (2017).

\bibitem{silva} A.~Silva, Phys.~Rev.~Lett. {\bf 101}, 120603 (2008).  

\bibitem{campos} L.~Campos Venuti and P.~Zanardi, Phys.~Rev.~A {\bf 81}, 022113 (2010).

\bibitem{keller} T.~Keller and T.~Fogarty, Phys.~Rev.~A {\bf 94}, 063620 (2016).

%
%

\bibitem{talkner} P.~Talkner, E.~Lutz and P.~H\"{a}nggi, Phys.~Rev.~E {\bf 75}, 050102(R) (2007).

\bibitem{barbar} M.~N.~Barbar, {\it Phase Transition and Critical Phenomena}, edited by C.~Domb and J.~L.~Lebowitz, (Academic Press, New York, 1983),
                 Vol.~8.

\bibitem{foot} In the LMG model, one is unable to define a correlation length, but one can find similar diverging behavior at the
               critical point in some other variables, such as the relaxation time which is inversely proportional to the energy gap.

\bibitem{campbell} S.~Campbell, Phys.~Rev.~B {\bf 94}, 184403 (2016).




%
%
%
%
%






\end{thebibliography}
\end{document}